\documentclass[]{tPHM2e}

\usepackage{graphicx,subfigure}
\usepackage[utf8]{inputenc}

\begin{document}

\title{Self-diffusion in a monatomic glassforming liquid embedded in the hyperbolic plane}

\author{F. Sausset$^{\rm a}$$^{\ast}$ and G. Tarjus$^{\rm a}$\thanks{$^\ast$Corresponding author. Email: sausset@lptmc.jussieu.fr
\vspace{6pt}}\\\vspace{6pt}  $^{\rm a}${\em{Laboratoire de Physique Théorique de la Matière Condensée, Université Pierre et Marie Curie-Paris 6, UMR CNRS 7600, 4 place Jussieu, 75252 Paris Cedex 05, France}}\\ }

\maketitle

\begin{abstract}
We study by Molecular Dynamics simulation the slowing down of particle motion in a two-dimensional monatomic model: a Lennard-Jones liquid on the hyperbolic plane. The negative curvature of the embedding space frustrates the long-range extension of the local hexagonal order. As a result, the liquid avoids crystallization and forms a glass. We show that, as temperature decreases, the single particle motion displays the canonical features seen in real glassforming liquids: the emergence of a ``plateau'' at intermediate times in the mean square displacement and a decoupling between the local relaxation time and the (hyperbolic) diffusion constant.
\end{abstract}

\section{Introduction}
Glass formation results from the slowing down of dynamics as a liquid is cooled to low enough temperature. Progress in a theoretical description of the phenomenon has been hampered by the shortage of an adequate and simple enough liquid model, which could play a role similar, say, to that of the Edwards-Anderson model in spin-glass theory \cite{Mezard:1987}. Motivated by earlier work based on the concept of ``frustration''\cite{Sadoc:1999,Nelson:2002,Tarjus:2005}, viewed as an incompatibility between extension of the locally preferred order in a liquid and tiling of the whole space, we have recently studied a monatomic two-dimensional glassforming model: the Lennard-Jones liquid embedded in the hyperbolic plane\cite{Sausset:2008}. The negative curvature characterizing the latter induces a frustration in the tendency of the system to order in a hexagonal structure (as it would do in the ``flat'' Euclidean space) and forces in an irreducible density of topological defects\cite{Rubinstein:1983}. As a result, long-range ordering is thwarted  and the two-dimensional monatomic liquid can now form a glass. Elsewhere, we have shown that (1) the ``fragility'' of the liquid, i.e. the extent to which the temperature dependence of the relaxation time deviates from a simple Arrhenius law and displays ``super-Arrhenius'' behavior, can be tuned at will by changing the curvature of the embedding space and (2) that the dynamics become spatially heterogeneous as temperature decreases\cite{Sausset:2008}. A potential interest of the model lies in its two-dimensional and monatomic character: this opens the way to theoretical developments based on the presence of well-identified point-like topological defects \cite{Nelson:1983b,Nelson:2002}.

In this short article, we consider another aspect of the dynamics of the Lennard-Jones liquid on the hyperbolic plane. We monitor the single particle (translational) motion as one lowers the temperature and approaches glass formation. More specifically, we study the properties of the long-time ``diffusive'' behavior and the decoupling of diffusion from local structural relaxation that one often associates with a breakdown of the Stokes-Einstein relation\cite{Sillescu:1998,Ediger:2000}.
One motivation for this study is to verify that the hyperbolic metric does not introduce spurious effects and that the present model indeed displays the phenomenology observed in real glassforming liquids.

\section{Model}

We have considered what seems to be the simplest microscopic liquid model that is subject to ``geometric frustration'': a monatomic liquid on the hyperbolic plane. Atoms interact through the standard Lennard-Jones pair potential $v(r) = 4\epsilon\left( (\sigma/r)^{12}-(\sigma/r)^{6} \right) $, where the distance $r$ is defined with the hyperbolic metric. The hyperbolic plane (denoted $H^2$) is a Riemannian surface of constant negative curvature ($K < 0$), whose metric in polar coordinates $(r, \phi)$ is given by

\begin{equation}
\mathrm{d}s^2= \mathrm{d}r^2 + \left(\frac{\sinh(\kappa r)}{\kappa} \right)^2 \mathrm{d}\phi^2,
\end{equation} 
where $\kappa$ is related to the Gaussian curvature $K$ of the plane by $K=-\kappa^2$.

The Euclidean counterpart of this system ($\kappa=0$) orders to a hexagonal ``crystal" via either a weakly first-order transition or a sequence of two nearby continuous transitions separated by an intermediate ``hexatic phase''\cite{Nelson:1983b,Strandburg:1988}. However, in nonzero curvature ($\kappa \neq 0$), the hexatic/hexagonal order that occurs in flat space can no longer tile the whole plane. If the curvature, as characterized by the dimensionless parameter $\kappa \sigma$, is small enough, the hexagonal order is still locally favored. Then, the local and global orders are incompatible and the system is (geometrically) frustrated\cite{Sadoc:1999,Nelson:2002,Tarjus:2005}; $\kappa^{-1}$ therefore plays the role of an intrinsic frustration length. The ordering transition that takes place in zero curvature at (or around) a temperature $T^*$ is avoided \cite{Kivelson:1995} for any nonzero curvature, no matter how small; the system remains in the liquid phase even below $T^*$ and may subsequently form a glass\cite{Sausset:2008}. By an abuse of language, we shall refer to the liquid below $T^*$ as being ``supercooled'' even though it does not compete with a crystalline phase when the curvature is nonzero. (As discussed in Refs. \cite{Tarjus:2005,Kivelson:1995}, one can associate $T^*$ in real three-dimensionnal glassforming liquids with a temperature that is empirically found to mark the onset of ``anomalous'' behavior characterized by super-Arrhenius temperature dependence of the relaxation time and non-exponential time dependance of the relaxation functions \cite{Sastry:1998}.)

We have performed a Molecular Dynamics simulation of the model. Details on the procedure that requires nontrivial extensions of the simulation method commonly used in Euclidean space are given in Ref. \cite{Sausset:2008}. The simulation is performed in the microcanonical ensemble $(N,V,E)$ and octagonal periodic boundary conditions are used\cite{Sausset:2007,Sausset:2008}. The control parameters are the temperature $T\epsilon^{-1}$, the density $\rho \sigma^2$, and the frustration associated with space curvature $\kappa \sigma$.

\section{Self-diffusion on the hyperbolic plane}

In this article, we focus on several aspects of the single particle motion. In a real liquid, it is well known that the long-time limit of such a motion is described by a diffusion process characterized by the (self-)diffusion coefficient $D$. Standard calculations show that the mean square displacement of the particles goes asymptotically as predicted by the Einstein relation: $\left\langle r(t)^2\right\rangle \sim 4 Dt$, where $r(t)=|\mathbf{r}(t)-\mathbf{r}(0)|$ is the absolute value of the displacement of a given particle during the time $t$. In a curved space, this relation does not hold anymore and one may wonder what is the generalization of the diffusive limit. To answer this point, we first take a little mathematical detour for considering the Brownian motion in the hyperbolic plane.

The Brownian motion in the hyperbolic plane has recently been studied\cite{Comtet:1996,Monthus:1996}. It is based on 
a Fokker-Planck equation describing the time evolution of the probability density $P (r, t)$ associated with finding the particle at a distance $r$ of its starting point after a time $t$:
\begin{equation}
\label{eq:FokkerPlanck}
\frac{\partial P (r, t)}{\partial t} = D \frac{\partial}{\partial r} \left[
\frac{\partial P (r, t)}{\partial r} - \kappa \coth (\kappa r) P (r, t)
\right],
\end{equation}
whose solution is
\begin{equation}
\label{eq:densityProbability}
P (r, t) = \frac{e^{- \frac{Dt}{4 \kappa^{- 2}}}}{2 \sqrt{2 \pi}
(Dt)^{\frac{3}{2}}}\, \sinh (\kappa r) \int_r^{\infty} \mathrm{d} y \frac{y \,e^{-
\frac{y^2}{4 Dt}}}{\sqrt{\cosh (\kappa y) - \cosh (\kappa r)}},
\end{equation}
where $D$ is the (hyperbolic) diffusion coefficient. Equation \eqref{eq:densityProbability} allows one to calculate the time dependence of the mean square displacement $\left\langle r^2(t)\right\rangle$ and the mean (absolute) displacement $\left\langle r(t)\right\rangle$. An asymptotic analysis then shows that when the distance travelled $r \gg\kappa^{-1}$, a ballistic diffusion is obtained with $\left\langle r^2(t)\right\rangle \sim (Dt)^2$ and $\left\langle r(t)\right\rangle \sim Dt$, whereas when $r \ll\kappa^{-1}$, ordinary (Euclidean-like) diffusion is found with $\left\langle r^2(t)\right\rangle \sim Dt$ and $\left\langle r(t)\right\rangle \sim (Dt)^{1/2}$. (Note that, of course, it is the same diffusion coefficient $D$ that appears in these two regimes.)

The working assumption we make to generalize the treatment of liquids in ordinary Euclidean space is that the single particle motion in the liquid embedded in the hyperbolic plane converges at long times to a process described by the above Brownian motion.

In practice, and to increase the domain over which one can use a simple fitting formula to extract the self-diffusion coefficient $D$ from the single particle motion, it is convenient to devise a function of the (absolute) displacement, $f(r(t))$, that behaves similarly in the intermediate Euclidean-like diffusive regime and in the asymptotic ballistic diffusive regime, i.e. such that $\left\langle f(r)\right\rangle \sim D\,t$ over a wide time domain. An expression of $f$ can be obtained by using the Fokker-Planck equation \eqref{eq:FokkerPlanck} and it reads

\begin{equation}
\label{eq:f(r)}
f(r(t))=\kappa^{-2}\ln\left( \frac{1+\cosh(\kappa r(t))}{2}  \right)\ .
\end{equation}

\begin{figure}[tbp]
\begin{center}
	\includegraphics[width=10cm]{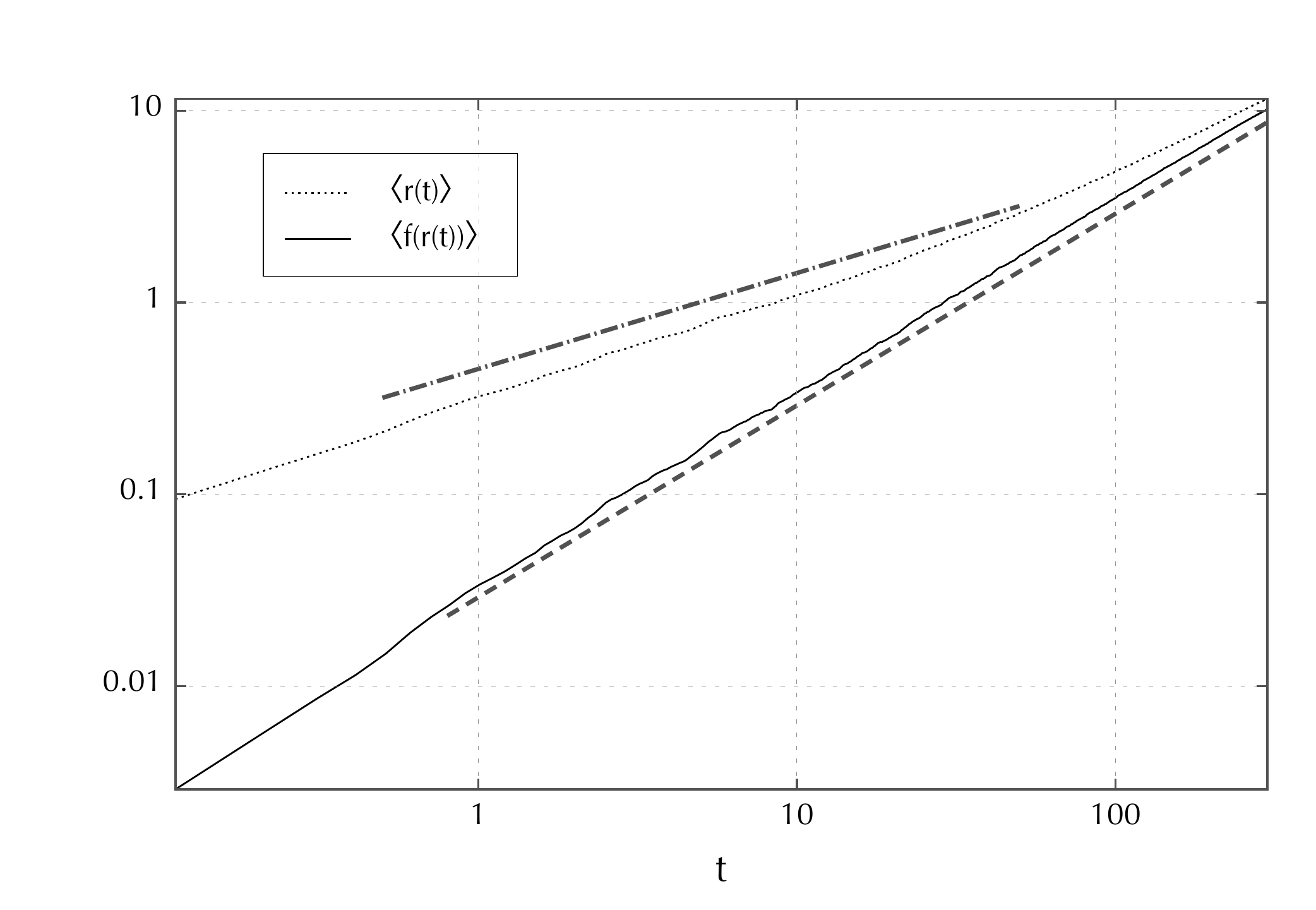}
\caption{Log-log plot of the mean absolute displacement $\left\langle r(t)\right\rangle$ in units of $\kappa^{-1}$ and of $\left\langle f(r(t))\right\rangle$ (see Eq. \eqref{eq:f(r)}). The dashed line has a slope equal to $1$ and the dotted-dashed one a slope of $1/2$. $\left\langle f(r(t))\right\rangle$ has a roughly linear time dependence at all times, which corresponds to a diffusive regime in the hyperbolic sense. The parameters are $\rho \sigma^2=0.91$, $T=2.17$, and $\kappa\sigma=0.2$. The system with octagonal periodic boundary conditions comprises 287 atoms.}
\label{fig:check}
\end{center}
\end{figure}

One easily checks that indeed $\left\langle f(r)\right\rangle = D\,t$ for a Brownian particle. Thus, by calculating $\left\langle f(r(t))\right\rangle=(1/N)\sum_{i=1}^N f(r_i(t))$, where the sum is over all the atoms of the system, and comparing it with a linear time dependence, one can verify that the atomic motion in the Lennard-Jones liquid becomes diffusive at long enough times and obtain the coefficient of self-diffusion $D$ without worrying about the crossover between ordinary and ballistic diffusion. That it is actually the case is illustrated on the model with frustration parameter $\kappa\sigma=0.2$, at the density $\rho \sigma^2=0.91$, and at a fairly high temperature ($T=2.17$, much above the ``avoided transition'' temperature $T^*\simeq 1.3$) so that the atoms move distances much larger than $\kappa^{-1}$ in the simulation time. The  results can be seen in figure \ref{fig:check}, where the mean absolute displacement $\left\langle r(t)\right\rangle$ and the associated $\left\langle f(r(t))\right\rangle$ are plotted versus time. One can see that, as expected, $\left\langle r(t)\right\rangle$ exhibits two different regimes in the diffusion limit: one following ordinary diffusion for $r \lesssim \kappa^{-1}$ and one following ballistic diffusion for $r\gg \kappa^{-1}$; $\kappa^{-1}$ thus sets the scale at which the crossover between the two regimes takes place. The figure also illustrates that it is easier to extract the self-diffusion coefficient $D$ from fitting $\left\langle f(r(t))\right\rangle$, which behaves as $Dt$ over a very wide domain.

\section{Single particle motion in the supercooled liquid}

In the ``supercooled'' liquid (i.e. the liquid below $T^*$, see above), one observes a dramatic slowing down of the dynamics. As shown in Ref.  \cite{Sausset:2008}, the superArrhenius character and the associated fragility increase as one decreases the frustration $\kappa\sigma$. Here, we focus on a moderately small frustration, $\kappa\sigma=0.1$, and on two features of the single particle dynamics at sufficiently low temperatures:  the emergence of a ``plateau" in the mean (absolute) displacement of the atoms and the decoupling of the diffusion and the local structural relaxation.

\subsection{Plateau in the mean (absolute) displacement}

\begin{figure}[tbp]
\begin{center}
	\includegraphics[width=10cm]{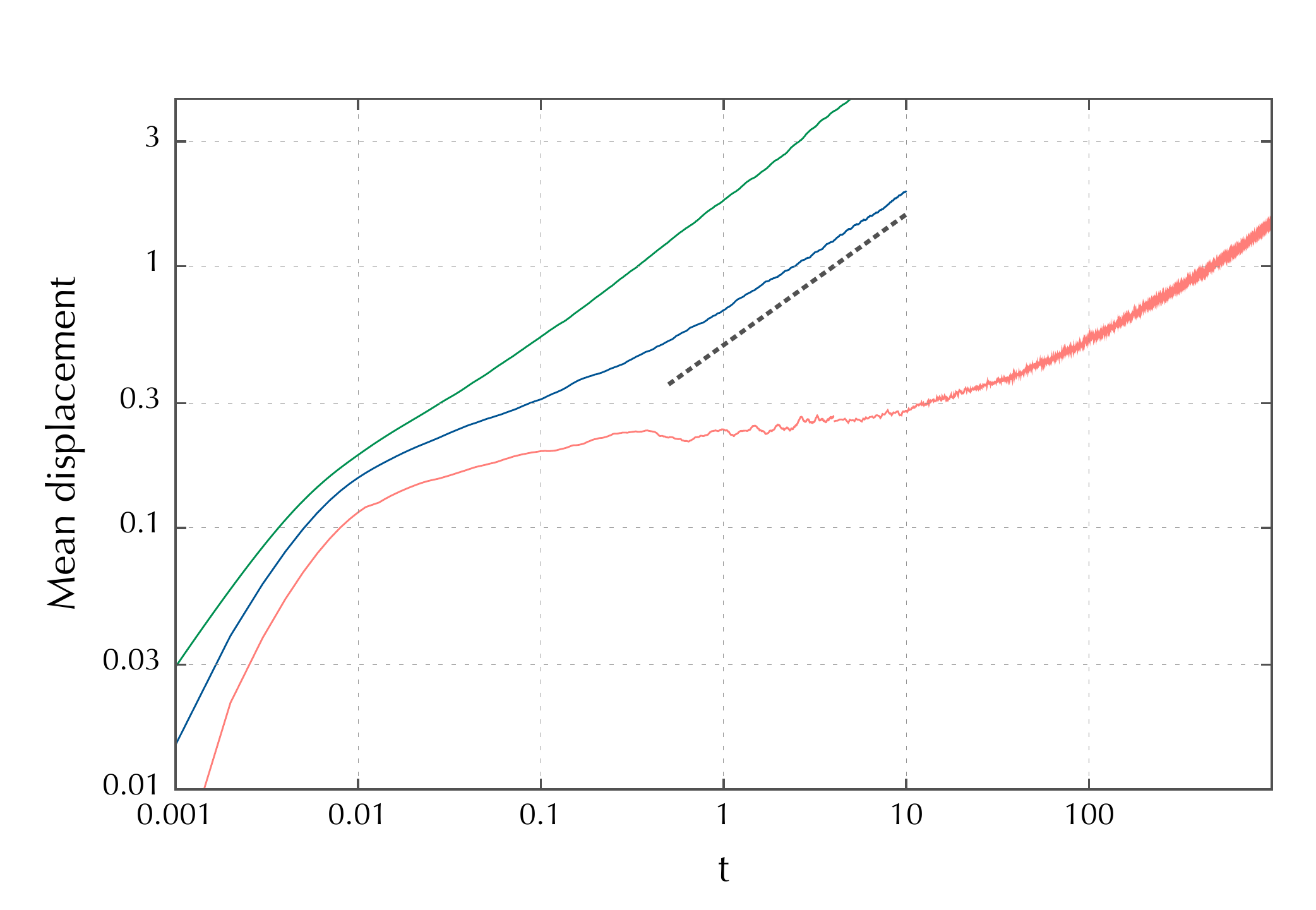}
\caption{Log-log plot of the mean absolute displacement $\left\langle r(t)\right\rangle$ of the atoms in units of the atomic diameter  $\sigma$. For this case, $\kappa\sigma=0.1$, which means that when $\left\langle r\right\rangle \sim \sigma$, it is only a tenth of $\kappa^{-1}$. Three temperatures are represented: from top to bottom, $T=1.53$, $T=0.96$, and $T=0.47$, with the avoided transition temperature being at $T^*\simeq1.3$. As $T$ decreases, a ``plateau" emerges at intermediate times and it is well developed at the lowest temperature. At longer times, a diffusive behavior is found, as shown by the dashed line whose slope is equal to $1/2$.}
\label{fig:plateau}
\end{center}
\end{figure}

We now compute the mean absolute displacement  $\left\langle r(t)\right\rangle$ of the atoms in a system of density $\rho\sigma^2=0.91$ for the liquid phase both above and below the avoided transition temperature $T^*\simeq 1.3$ at which ordering occurs in the Euclidean plane: see Fig. \ref{fig:plateau}. At high temperature above $T^*$,  $\left\langle r(t)\right\rangle$ is very close to its counterpart in the Euclidean plane in the domain covered, i.e. for $\left\langle r\right\rangle$ up to a few atomic diameters. On such distances that are significantly smaller than the intrinsic frustration length $\kappa^{-1}$ (recall that $\kappa^{-1}= 10\sigma$ here), the dynamics is dominated by local relaxation processes unaffected by the curvature. At short times, $\left\langle r(t)\right\rangle$ is ballistic due to the Newtonian dynamics of the atoms and it crosses over to a diffusive behavior at longer times. When decreasing the temperature and entering the supercooled liquid regime, a plateau appears in $\left\langle r(t)\right\rangle$ at intermediate times and for a displacement of a fraction of $\sigma$. This plateau becomes more pronounced as $T$ is lowered and for the lowest temperature $\left\langle r(t)\right\rangle$ stays roughly constant for about two decades of time. This phenomenon is usually described as a ``cage effect": particles are trapped by their neighbors (which are themselves trapped) and they can only rattle before being able to escape the cage and change their neighborhood\cite{Gotze:1992}. At long times, the atomic motion becomes diffusive. All of this is very similar to what is found in simulations and experiments on $3$-dimensional glassforming liquids\cite{Kob:1995,Weeks:2000}.

\subsection{Decoupling of diffusion and local structural relaxation}

Another intriguing experimental observation found in supercooled liquids is the decoupling of the characteristic time scales measured by different probes (translational diffusion on one hand, viscosity and $\alpha$-relaxation time on the other hand) and the associated breakdown of the Stokes-Einstein relation\cite{Sillescu:1998,Ediger:2000}. In ordinary liquids (above the melting point), a homogeneous description is found to apply down to molecular scale and the hydrodynamic-based Stokes-Einstein relation, $\frac{D\,\eta}{T}= const.$ with $D$ the self-diffusion coefficient and $\eta$ the viscosity, is well obeyed\cite{Hansen:1986}. However, experimental measurements have shown that this relation does not hold anymore in the supercooled liquid regime, where the ratio $\frac{D\,\eta}{T}$ increases with decreasing temperature by up to several orders of magnitude\cite{Sillescu:1998,Ediger:2000,Swallen:2003}. On the other hand, the $\alpha$-relaxation time $\tau_\alpha$ obtained from rather local measurements, i.e. probes involving particle displacements over a small, molecular scale, such as rotational and structural relaxation measurements, follow rather well the temperature dependence of the viscosity and therefore also decouple from the self-diffusion process that implies particle translational motion over large, super-molecular distances.

\begin{figure}[tbp]
\begin{center}
	\includegraphics[width=10cm]{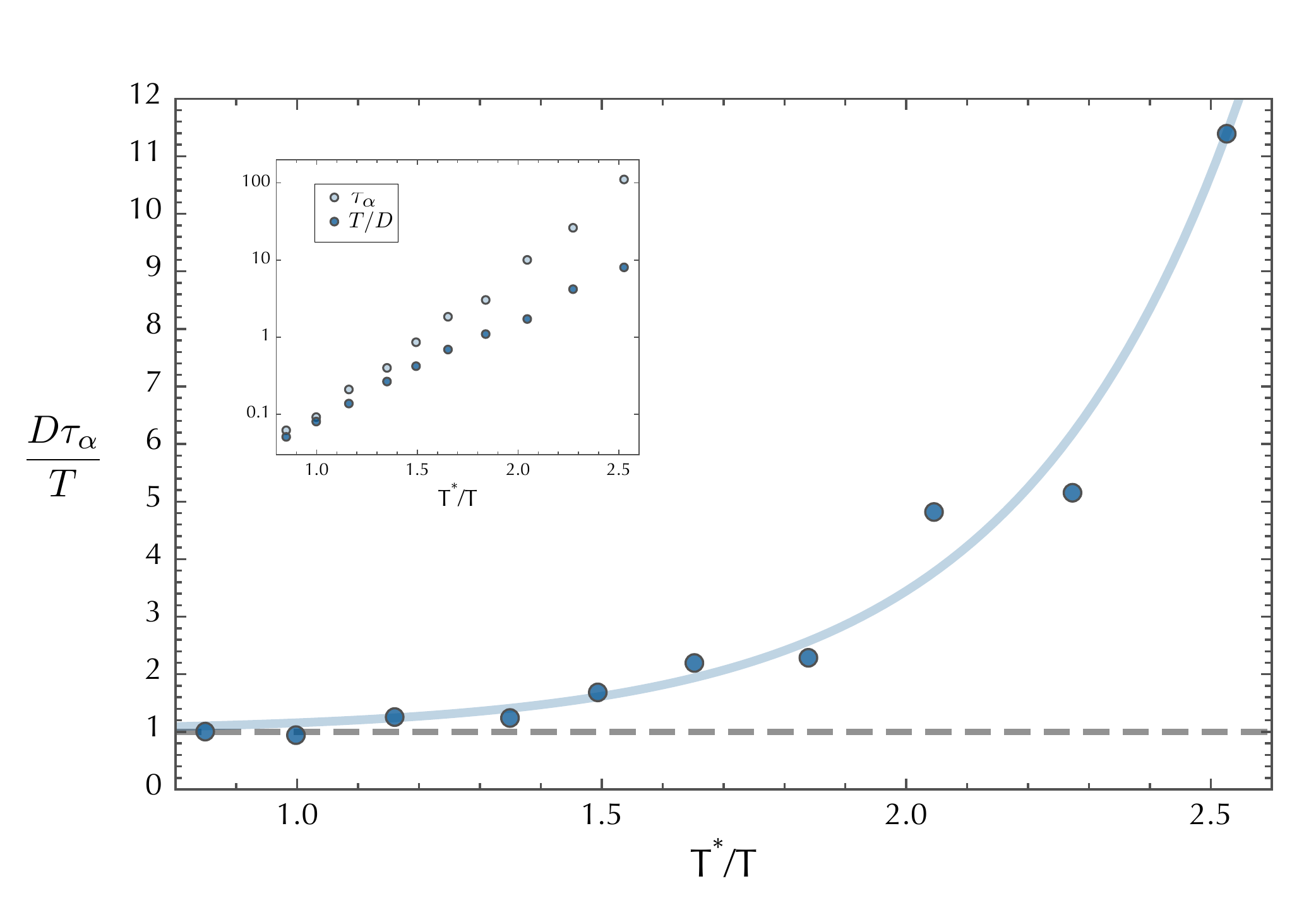}
\caption{Decoupling ratio $\frac{D\,\tau_\alpha}{T}$ between the diffusion and the alpha-relaxation versus the inverse temperature. The ratio is normalized to 1 at high temperature and the temperature is scaled by $T^*$.  In the inset we separately display $\frac{D}{T}$ and $\tau_\alpha$ in an Arrhenius plot; the scales on the $y$-axis are shifted in order to bring the two sets of data together at high temperature. }
\label{fig:ratio}
\end{center}
\end{figure}

To study such a potential decoupling in our model, we have monitored both the self-diffusion coefficient $D$ (see above) and the $\alpha$-relaxation time $\tau_\alpha$ obtained from the hyperbolic generalization of the self-intermediate scattering function $F_s(k,t)$, when studied for a wavevector $k\simeq \sigma^{-1}$ close to the first maximum of the static structure factor\cite{Sausset:2008}. The temperature dependences of $T/D$ and $\tau_\alpha$ are displayed in the inset of Fig. \ref{fig:ratio} in an Arrhenius plot. As can be seen from the plot, the deviation from Arrhenius behavior is not very marked for this curvature ($\kappa\sigma=0.1$) and this density ($\rho\sigma^2=0.91$); the corresponding glassforming liquid is rather ``strong'' (or in a terminology that is somewhat more appropriate for atoms interacting via weak nondirectional pair potentials, has a small fragility\cite{Ferrer:1999}). Nonetheless, there is a significant increase, by an order of magnitude, of the analog of the  Stokes-Einstein ratio, i.e. $\frac{D\,\tau_\alpha}{T}$, as temperature decreases in the supercooled regime: see Fig. \ref{fig:ratio}. One expects that the magnitude of the decoupling would increase as the fragility of the liquid increases. This can be investigated in our model by decreasing the curvature $\kappa\sigma$ of the embedding space, and this work is in progress.

\section{Conclusion}

We have illustrated that the frustration-based microscopic model of a two-dimensional monatomic Lennard-Jones liquid embedded in the hyperbolic plane displays many features of the phenomenology of supercooled liquids (see also \cite{Sausset:2008}). In this short article, we have focused on the single particle motion, showing that the long-time behavior is described by a hyperbolic generalization of the diffusion process and devising an efficient way to extract the self-diffusion coefficient. The two canonical features of glassforming liquids as temperature is lowered that we have retrieved are the emergence of a plateau in the time dependence of the  mean (absolute) displacement of the atoms and a decoupling of the diffusion and the $\alpha$-relaxation. One main advantage of the model is that by changing the space curvature one varies the fragility of the system so that one can envisage the phenomenology of glassforming liquids at constant molecular parameters but varying fragility. In addition, one expects that the model may lend itself to a theoretical treatment based on the consideration of the point-like topological defects that are characteristic of two-dimensional, one-component systems. 

\bibliographystyle{tPHM} 
\bibliography{Andalo} 

\end{document}